\documentclass[twocolumn,prd,showpacs,amsmath,amssymb]{revtex4}

\begin{document}
\title{ Quantum fluctuations of braneworld background \\ versus the TeV fundamental scale}

\author{Michael~Maziashvili}
\email{maziashvili@hepi.edu.ge}\affiliation{Department of Physics,
Tbilisi
State University, 3 Chavchavadze Ave., Tbilisi 0128, Georgia \\
Institute of High Energy Physics and Informatization, 9 University
Str., Tbilisi 0186, Georgia }

\begin{abstract}
Quantum fluctuations of gravitational field with respect to the ADD braneworld
background turns out to be large enough spoiling the TeV scale physics as
well. In view of this observation it becomes important to look for some
protecting mechanism. The most natural such mechanism in the case of
braneworld model with compactified extra dimensions can be the shape moduli of
extra space. Quantum fluctuations of braneworld  
background affect significantly the production as well as evaporation of mini
black holes studied very actively during the las few years. The enormous
amplification of quantum fluctuations of the background space-time due to
lowering the fundamental scale to the TeV range, is
characteristic to other phenomenological braneworld models as well and
correspondingly calls for attention.       
\end{abstract}



\maketitle

The idea of large extra dimensions proposed by Arkani-Hamed, Dimopoulos and
Dvali \cite{ADD} attracted considerable continuing interest over the past
several years. If the fundamental scale of gravity is indeed in the TeV range,
one expects that extra dimensions should start to show up in accelerator and
astrophysical experiments at energies approaching this scale \cite{ADD1}. The lowering of
fundamental gravity scale to the TeV range opens up new possibilities for the
black hole physics as well. Namely, it gives rise to a intriguing possibility
of TeV scale black hole production at the near-future accelerators as well as
in high energy cosmic ray experiments \cite{DLGTFS}. In this paper we notice
that in phenomenological braneworld models quantum fluctuations of background 
space-time become discouragingly enlarged due to setting the fundamental scale
much below the Planck one. For definiteness let us indicate that the quantum fluctuations we deal with have different character relative
to the standard perturbation field with respect to the background metric for which the theoretical framework describing the
dynamics is well established. The point is that there is an unavoidable quantum uncertainty in
space-time measurement that naturally translates into the metric fluctuations
\cite{SW}. In what follows we will be concerned with uncertainties due to
quantum fluctuations of the background metric, in other words fluctuations
that are introduced when a measurement is performed.

 First let us briefly recapitulate the results
concerning the issue of quantum fluctuations of the background Minkovskian
space. In what follows we adopt the units $\hbar=c=1$. The minimal uncertainty in length
measurement has the following form \cite{Ma}
\begin{equation}\label{uncleng}\delta l_{min}=2^{{3+2n\over 3+n}}A^{{1\over
3+n}}\left (l_F^{2+n}l \right)^{{1\over 3+n}}+2^nAl_F^{2+n}l^{-(1+n)}~,
\end{equation} where $n$ denotes the number of extra dimensions, $l_F$
stands for fundamental length scale and \[ A={8\Gamma\left({3+n\over
2}\right)\over  (2+n)\pi^{{n+1\over
2}}}~.\] (For $n=0$ the fundamental length equals
the Planck one $l_{Pl}\sim 10^{-33}$cm). The Eq.(\ref{uncleng}) exhibits the minimal
observable length comparable to $\sim l_F$. The minimal uncertainty
in time measurement, $\delta t_{min}$, can be obtained from
Eq.(\ref{uncleng}) simply by replacing $l_F$ and $l$ with $t_F$
and $t$ respectively. Since the first term in Eq.(\ref{uncleng}) is much
greater than the second one for $l\gg l_F$ and they become comparable at $l\sim l_F$, one can say with no loss of generality
that for $l\gtrsim l_F$ our precision of space-time measurement is
limited by the measurement process itself such that \begin{equation}\label{stuncert}\delta l\sim
l_F^{\alpha}l^{1-\alpha}~,~~~~~~~~\delta t\sim
t_F^{\alpha}t^{1-\alpha}~, \end{equation} where $\alpha=(2+n)/(3+n)$.
Correspondingly for background metric fluctuations over a region with linear
size $l$ one gets  \[ \delta g _{\mu\nu}   \sim
\left({l_F\over l}\right)^{\alpha}~.\]
The metric fluctuations, $\delta g _{\mu\nu} $, result
in the energy-momentum uncertainties as well. Namely, the particle with momentum $p$
has the wavelength $\lambda$ where $\lambda\sim p^{-1}$ and from the above
relation for length uncertainty one simply gets \begin{equation}\label{emuncert}\delta p\sim
{p^{1+\alpha}\over m_F^{\alpha}}~,~~~~~~\delta E\sim
{(E^2-m^2)^{{2+\alpha\over2}}\over E\, m_F^{\alpha}}~.\end{equation} 

The consideration of space-time uncertainties can be immediately generalized to the braneworld
scenario \cite{Ma, Ma1}.
Let us consider the ADD braneworld model with extra dimensions running
from $0$ to $2\pi L$ where the points $0$ and $2\pi L$ are identified \cite{ADD}. Without going
into much details let us merely recall a few basic features
relevant for our consideration. There is a low fundamental scale,
 $m_F\sim$TeV, the standard model particles are
localized on the brane while the gravity is allowed to propagate
throughout the higher dimensional space. Loosely speaking there is a length scale
$L$ beneath of which the gravitational
interaction has the higher dimensional form due to contribution from KK gravitons whereas beyond this
scale we have the standard four-dimensional law produced by the zero mode of
KK spectrum

\begin{equation}\label{potential}V(r)=\left\{\begin{array}{ll} l_F^{2+n}m/ r^{1+n}~, & \mbox{for}~~ r\leq L~,\\\\
l_{Pl}^2m/ r~, & \mbox{for}~~ r>L~.
\end{array}\right.\end{equation} 
The discussion of minimal length
uncertainty goes as follows \cite{Ma, Ma1}. The distance measurement between
two points is performed by sending the light signal from the clock to the
mirror situated at those points respectively. As it is shown in \cite{SW} by
choosing the optimal size of the clock being of the order of $\sim\sqrt{l/m}$,
the total uncertainty in distance measurement is reduced to the size of the
clock. This quantity is bounded from below by the gravitational radius of the
clock determining the minimal unavoidable uncertainty in length measurement
\cite{SW}. In order to take into account the uncertainties contributed to the
measurement both by the clock and the mirror, one can use the \emph{gedanken}
experiment proposed in \cite{Ma}. The brane localized observer
using a clock with $r_s < L$ and measuring a distance $l<L$ finds the
Eq.(\ref{uncleng}) for minimal length uncertainty. In the case if $l$ is
greater than $L$ but the size of clock is still less than $L$ the second term
in Eq.(\ref{uncleng}) appears with $n=0$ and $l_F\rightarrow l_{Pl}$. If the size of clock is greater than
$L$ one gets the Eq.(\ref{uncleng}) with $n=0$ and $l_F\rightarrow l_{Pl}$, i.e., the pure four-dimensional
result. Strictly speaking the transition of higher-dimensional gravity from
the region $r\ll L$ to the four-dimensional law for $r\gg L$ is more
complicated near the transition scale $\sim L$ than it is schematically described in Eq.(\ref{potential}), but
for the purposes of this paper it is less significant. Let us notice that in
the case of brane there are additional uncertainties in space-time
measurements caused by the brane width \cite{Ma2}.     

For favorable clock mass providing the
minimal uncertainty in measuring the distance $l$ one finds \cite{Ma1} 
\begin{equation}\label{clmass}m\sim l^{{1+n\over 3+n}}l_F^{-{4+2n\over 3+n}}~,
\end{equation} and correspondingly its gravitational radius takes
the form
 \[r_g\sim l^{{1\over
3+n}}l_F^{{2+n\over 3+n}}~.\] Hence, for the region with linear size $l$ beneath the scale \[h \sim
L^{3+n}l_F^{-(2+n)}~,\] the uncertainties in space-time intervals are given by
Eq.(\ref{stuncert}). So that if the particle probes the length scale less
than $h$, i.e. $p^{-1}\lesssim h$ its energy-momentum uncertainty takes the
form given by Eq.(\ref{emuncert}).  Postulating the
TeV fundamental scale in ADD model one finds $L\sim 10^{30/n-17}$cm
\cite{ADD}. Correspondingly one gets $n=2~,~h\sim 10^{54}$cm; $n=3~,~h\sim
10^{22}$cm; $n=4~,~h\sim
10^{30}$cm; $n=5~,~h\sim
10^{24}$cm; $n=6~,~h\sim
10^{20}$cm.

From Eq.(\ref{emuncert}) one sees that for the particle with the
mass $m\ll m_F$ and energy $E\sim m_F$, the uncertainty in energy becomes
comparable to the energy itself. So that the quantum fluctuations of space-time become unacceptable amplified
in this case even for the TeV scale physics. For ultra high energy cosmic rays with $E\sim
10^{8}$TeV the uncertainty in energy becomes greater than $10^{13}$TeV. Hence,
the ultra high energy cosmic rays put the restriction on the fundamental scale
$m_F\gtrsim 10^{8}$TeV. The discouragingly amplified uncertainties in energy-momentum can be seen in a more simple
way as well. The brane localized particle with momentum grater than $L^{-1}$,
($L^{-1}\ll$TeV ) probes the length scale beneath $L$, the gravitational law
for which is higher-dimensional and therefore in this case the
Eq.(\ref{emuncert}) is directly applicable. As it is briefly emphasized in
\cite{Ma}, in order to see the unacceptable
magnification of these fluctuations one can consider them in light of stellar
interferometry observations as well \cite{LH, RTG}.

Now let us consider the effect of space-time fluctuations on the black hole
production and subsequent evaporation. For the emission temperature and entropy of
higher-dimensional black hole one finds \cite{DLGTFS} \[T={n+1\over 4\pi
r_g}~,~~~~~~~~S={Ar_g^{2+n}\over 4l_F^{2+n}}~.\] Due
 to length uncertainty, Eq.(\ref{stuncert}), the horizon of the black
hole undergoes the quantum fluctuations (at least) of the order $\delta r_g\sim
l_F^{\alpha}r_g^{1-\alpha}$ resulting thereby in the fluctuations of black
hole thermodynamics \[\delta T\sim {1\over r_g-\delta r_g}-{1\over
r_g}~,~~\delta S\sim {(r_g+\delta r_g)^{2+n}-r_g^{2+n}\over l_F^{2+n}}~.\] The cross-section of the black hole production is
proportional to the square of the horizon area \cite{DLGTFS}  $\sigma\sim r_g^2$ and
therefore undergoes fluctuations of the order \[\delta \sigma \sim (r_g+\delta
r_g)^2-r_g^2~.\] Near the fundamental scale the fluctuations $\delta
T,~\delta S,~\delta\sigma $,  become of the
same order as $T,~S,~\sigma$, and therefore
does not allow one to say something definitely about the production and
subsequent evaporation of mini black holes. When the brane localized black hole has
evaporated down to the fundamental length size, the standard thermodynamic
theory of the black hole is no longer applicable, as space-time is subject to
violent quantum fluctuations on this scale.                

In general when $r_g\gg \delta r_g$, i.e., $r_g^{\alpha}\gg l_F^{\alpha}$, the
temperature, entropy and production cross section of the black hole undergo
fluctuations given by  \[\delta T\sim {l_F^{\alpha}\over r_g^{1+\alpha}}~,~~\delta S\sim
\left({r_g\over l_F}\right)^{2+n-\alpha}~,~~\delta\sigma \sim l_F^{\alpha}r_g^{2-\alpha}~.\] 

We see that in ADD braneworld model with TeV fundamental scale
the foamy structure of space-time shows up at energies approaching this scale
threatening therefore the TeV scale physics.
One can protect the model from these amplified fluctuations by increasing accordingly the mass gap between the zero and first
excited KK graviton modes while keeping the fundamental scale into the TeV
range. As it is shown in paper \cite{Di} such result can be obtained by means
of the compactification geometry associated with extra dimensions. Namely, the degree
to which the Planck scale may be lowered depends on the volume of compactified
dimensions. However, the shape moduli of extra space can have significant
effect on the corresponding KK spectrum. For certain shape moduli it is
possible to maintain the ratio between the higher-dimensional fundamental
scale and the Planck one while simultaneously increasing the KK graviton mass
gap by an arbitrary large factor. Such KK masses completely avoid direct
laboratory bounds from precision tests of non-Newtonian gravity and alleviate
(and perhaps even eliminate) many of the bounds that constrain theories with
large extra dimensions considered in \cite{ADD1} and multitudinous subsiquent
papers. In most previous discussions of large extra dimensions little
attention has been paid to the implications of shape moduli. But the above
consideration makes it mandatory to protect the model from quantum
fluctuations of background space-time and shape moduli seems to be one of the most
natural protecting mechanisms. The above discussion illustrates a
general feature of the quantum fluctuations of braneworld background
irrespectively of the concrete model. Unfortunately, we don't know what can be
the possible protecting mechanism from these fluctuations in the case of braneworld models with TeV fundamental
scale and without compactified extra dimensions.

\begin{acknowledgments}
The author is greatly indebted to David Langlois for cordial hospitality at
APC (Astroparticule et Cosmologie, CNRS, Universit\'e Paris 7) and IAP
(Institut d'Astrophysique de Paris), where this work was done. Thanks are due
to M.~Bucher, C.~Deffayet and M.~Minamitsuji for useful conversations. The work was supported by
the \emph{INTAS Fellowship for Young Scientists}, the
\emph{Georgian President Fellowship for Young Scientists} and the
grant \emph{FEL. REG. $980767$}.
\end{acknowledgments}

\end{document}